\documentclass[fleqn,usenatbib]{mnras}

\usepackage{newtxtext,newtxmath}
\usepackage[T1]{fontenc}
\usepackage{ae,aecompl}
\usepackage{graphicx}	
\usepackage{amsmath}	
\usepackage{amssymb}	
\usepackage{marvosym}
\usepackage{hyperref}

\newcommand{\beq}[1]{\begin{equation}\label{#1}}
\newcommand{\eeq}{\end{equation}}
\newcommand{\beqn}[1]{\begin{eqnarray}\label{#1}}
\newcommand{\eeqn}{\end{eqnarray}}
\newcommand{\sub}[1]{_\mathrm{#1}}

\newcommand{\Op}{\Omega\sub{p}}
\newcommand{\Os}{\Omega\sub{\star}}
\newcommand{\Ocri}{\Omega\sub{c}}
\newcommand{\Rc}{R\sub{c}}
\newcommand{\Mc}{M\sub{c}}
\newcommand{\rhoe}{\rho\sub{e}}
\newcommand{\rhoc}{\rho\sub{c}}
\newcommand{\rhop}{\rho\sub{p}}

\newcommand{\Oini}{\Omega\sub{p,ini}}
\newcommand{\nini}{n\sub{p,ini}}

\newcommand{\Rp}{R\sub{p}}

\newcommand{\Mp}{M\sub{p}}

\newcommand{\Mjup}{\mathrm{M}\sub{J}}
\newcommand{\Rjup}{\mathrm{R}\sub{J}}
\newcommand{\Msun}{\mathrm{M}_{\sun}}
\newcommand{\Rsun}{\mathrm{R}_{\sun}}
\newcommand{\Mearth}{\mathrm{M}_{\oplus}}

\newcommand{\Mstar}{M\sub{\star}}
\newcommand{\Rstar}{R\sub{\star}}

\newcommand{\apos}{a\sub{p}}

\newcommand{\npp}{n\sub{p}}

\newcommand{\Kpp}{k\sub{2,p}}
\newcommand{\Qp}{Q\sub{p}}
\newcommand{\Kss}{k\sub{2,\star}}
\newcommand{\Qs}{Q_\star}
\newcommand{\koQ}{\Kpp/\Qp}

\newcommand{\Der}{\mathrm{d}}
\newcommand{\AU}{\mathrm{au}}
\newcommand{\porb}{P\sub{orb}}
\newcommand{\prot}{P\sub{\star}}

\newcommand{\Stide}{S\sub{tide}}
\newcommand{\epp}{\varepsilon\sub{p}}
\newcommand{\eps}{\varepsilon\sub{\star}}
\newcommand{\gyrp}{\zeta\sub{p}}
\newcommand{\gyrs}{\zeta\sub{\star}}
\newcommand{\omp}{\Dot{\omega}\sub{p}}
\newcommand{\oms}{\Dot{\omega}_\star}

\newcommand{\alphap}{\alpha\sub{p}}
\newcommand{\betap}{\beta\sub{p}}
\newcommand{\gammap}{\gamma\sub{p}}

\newcommand{\alphas}{\alpha\sub{\star}}
\newcommand{\betas}{\beta\sub{\star}}
\newcommand{\gammas}{\gamma\sub{\star}}

\newcommand{\hl}[1]{\textcolor{black}{#1}}
\newcommand{\hll}[1]{\textcolor{black}{#1}}

\title[\hl{Orbital decay under evolving tides}]{Orbital decay of short-period gas giants under evolving tides}

\author[J. A. Alvarado-Montes and C. Garc\'ia-Carmona]{
Jaime A. Alvarado-Montes$^{1,2,3}$\thanks{E-mail: jaime-andres.alvarado-montes@hdr.mq.edu.au} and 
Carolina Garc\'ia-Carmona$^{3}$\thanks{E-mail: carolina.garcia8@udea.edu.co
}
\\
$^{1}$ Centre for Astronomy, Astrophysics and Astrophotonics, Macquarie University -- Sydney, NSW 2109, Australia.\\
$^{2}$ Department of Physics \& Astronomy, Macquarie University -- Sydney, NSW 2109, Australia.\\
$^{3}$ Solar, Earth, and Planetary Physics Group (SEAP), Instituto de F\'{\i}sica, FCEN, Universidad de Antioquia --\\
\hspace{0.2cm}Calle 70 No. 52-21, Medell\'in, Colombia.\\
}
\date{Accepted 2019 April 12. Received 2019 April 12; in original form 2019 January 18.}

\pubyear{2019}

\begin{document}
\label{firstpage}
\pagerange{\pageref{firstpage}--\pageref{lastpage}}
\maketitle

\begin{abstract}
The discovery of many giant planets in close-in orbits and the effect of planetary and stellar tides in their subsequent orbital decay have been extensively studied in the context of planetary formation and evolution theories. Planets orbiting close to their host stars undergo close encounters, atmospheric photoevaporation, orbital evolution, and tidal interactions. In many of these theoretical studies, it is assumed that the interior properties of gas giants remain static during orbital evolution. Here we present a model that allows for changes in the planetary radius as well as variations in the planetary and stellar dissipation parameters, caused by the planet's contraction and change of rotational rates from the strong tidal fields. In this semi-analytical model, giant planets experience a much slower tidal-induced circularization compared to models that do not consider these instantaneous changes. We predict that the eccentricity damping time-scale increases about an order of magnitude in the most extreme case for too inflated planets, large eccentricities, and when the planet's tidal properties are calculated according to its interior structural composition. This finding potentially has significant implications on interpreting the period-eccentricity distribution of known giant planets as it may naturally explain the large number of non-circularized, close period currently known. Additionally, this work may help to constrain some models of planetary interiors, and contribute to a better insight about how tides affect the orbital evolution of extrasolar systems.

\end{abstract}

\begin{keywords}
planets and satellites: dynamical evolution and stability -- planets and satellites: physical evolution -- planets and satellites: gaseous planets
\end{keywords}



\section{Introduction}
\label{sec:intro}

Since the first sub-stellar companion was discovered by \cite{Mayor1995}, we know that migration of giant planets is a common phenomenon in planetary systems (see e.g. \citealt{Armitage2010} and references therein). These giant planets could have been perturbed from their original distant orbits due to encounters with other planets \citep{BarnesR2006,Armitage2010,Rodriguez2011}, or even close stars \citep{Rasio1996,Marcy1997,Murray1998,Jackson2008a}. Also, tidal-induced interactions of a proto-planetary disk with a planet may induce a small planet's semimajor axis \citep*{Lin1996}, and in such a scenario the damping of high eccentricities is a natural consequence of planet-disk resonances for planets with masses of few times that of Jupiter \citep*{Goldreich1980,Artymowics1992,Papaloizou2001}, although sometimes the planet may acquire high eccentricities \citep{Goldreich2003}.

Many of these migrated Jupiter-like gas giants, a.k.a hot Jupiters, end in very close orbital distances, and although different processes might prevent further migration (e.g. the tidal interaction with their host star, or the disk depletion  \citealt{Lin1996}), they may also produce a high-eccentricity condition that can lead to circularization and tidal locking of the planetary orbit (i.e. synchronisation), as well as decay to even closer positions \citep*{Dobs2004,Ferraz2008,Jackson2008a}. 

\begin{figure}
    \centering
        \includegraphics[scale=0.53]{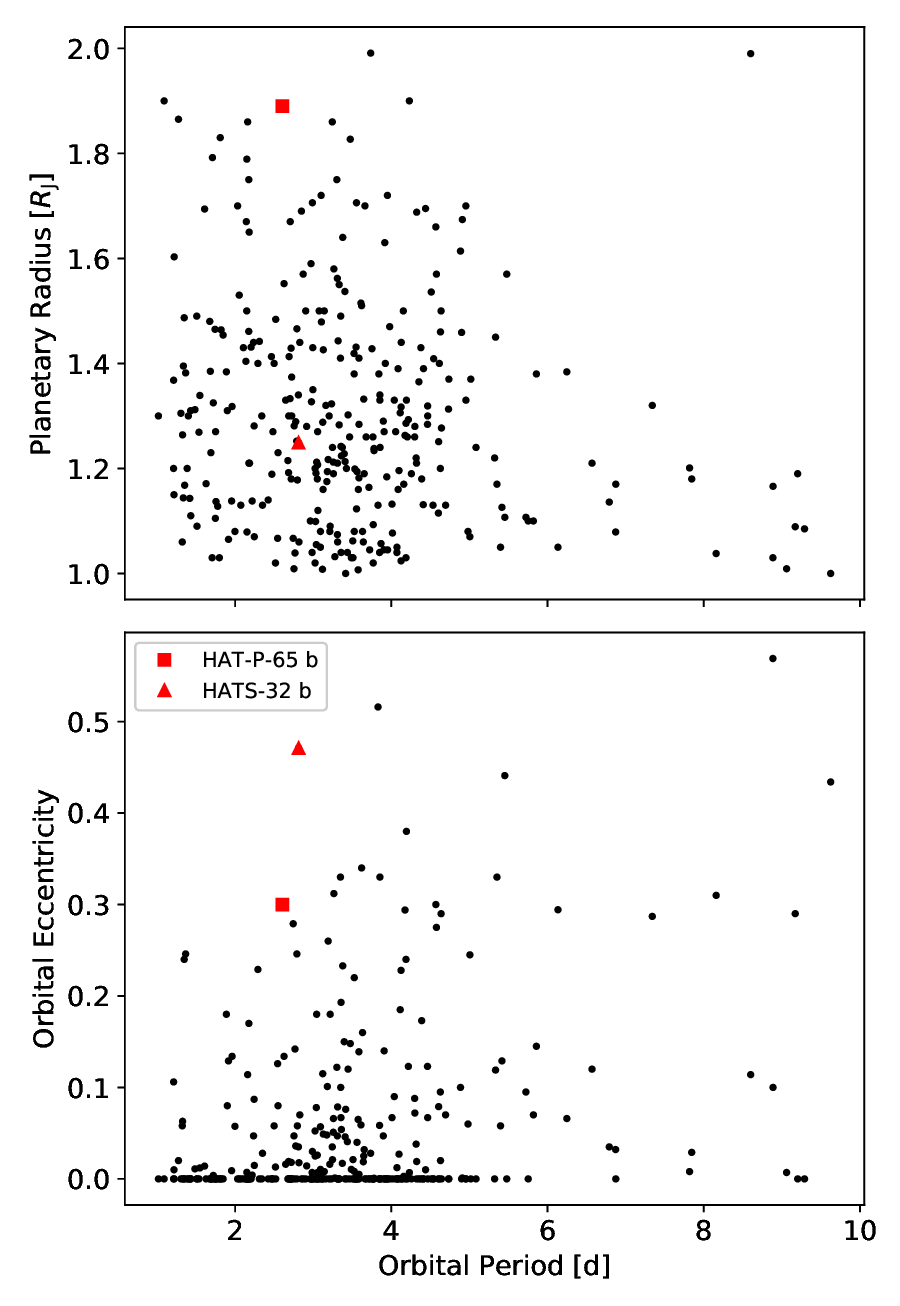}
    \caption{Jupiter-like exoplanet distributions of planetary radii (top panel) and eccentricities (bottom panel) respect to the orbital period. As  we can see there is an accumulation of eccentricities near (or equal) to zero, which implies that most close-in planets are already circularized. This was extracted and filtered from the database by \citet{exoeu}. The example planets for tidal evolution computed in Section \ref{sec:results}, namely HAT-P-65 b and HATS-32 b, are represented by red square and triangle, respectively. The data of this plot was downloaded on 22 March 2019.}
\label{fig:distrib}
\end{figure}

Observations reveal that $\sim250$ short-period exoplanets ($\porb<10$ d) have negligible orbital eccentricities (i.e. $e\sim0$), so regardless of their early dynamics they had to undergo an efficient damping mechanism whereby acquired those nearly circular orbits. Furthermore, there are more than 60 close-in planets with eccentricities from 0.1 to about 0.6 (see Fig. \ref{fig:distrib}), and from basic theoretical expectations (\hl{e.g.} \citealt{Rodriguez2010}) it is likely that those planets whose orbits are still eccentric ($e\ge0.1$) will somehow be affected by stellar/planetary tides and eventually circularize.  

In light of the above, star-planet tidal interactions are highly important in circularizing the orbits of short-period planets. Still, the strength of these interactions changes with the non-static structural composition of deformed bodies, which is not well constrained for planets and stars and is affected by the  system's angular momentum exchange. In other words, to model the orbital evolution of a close-in planet we must have into account how the stellar/planetary tidal-related properties are changing over time, because this is directly connected to the total dissipated energy in the star-planet system.

From the classical theory of tidal evolution, the dissipation properties of a body are usually represented with two `obscure' quantities that hinder our understanding of \hl{tides}. The first is the second-order tidal Love number $k_2$, which is a numerical coefficient determined by the tidal-effective rigidity of the deformed body, its radial bulk density, etc., and measures the response of a body to tidal stresses. The second is the tidal frequency-dependent $Q$ value, a dimensionless parameter that stands for: (1) the dissipation efficiency of tides in a distorted body which undergoes a forced oscillation, and (2) its tidal-related effects that play a key important role on the evolution of planetary systems, as shown by \cite{Goldreich1966} for a calibrated $Q$ in the Solar System. 

Both $k_2$ and $Q$ are not well constrained, so their coupled evolution and impact on planetary tidal decay and circularization is not well established. Also, how the tidal friction is conducted in exoplanets remains partially understood, and the underlying complex dissipative processes in solid (planets) or fluid (stars) layers of a body lack a proper understanding \citep{Zahn2008}. Even though, the dynamics of star-planet and planet-moon systems significantly depends on these unclear processes \citep*{Efroimsky2007,Auclair2014,Alvarado2017}. 

\hl{Various studies have proven that changes on these properties are quite important and have to be taken into account when computing the tidal evolution of close-in giant planets. For instance, \cite{Dobs2004} implemented a model which uses frequency-dependent changes in primary tidal components, e.g. effective $Q$. Other studies as \cite{Ferraz2008, Ferraz2015} consider tidal evolution tracks to take into account variations in the dissipation parameters, and use different values for the tidal dissipation reservoirs, the latter including the effect of magnetic braking. Also, \cite{Barker2009} compute different tidal functions directly related to the magnetic braking relaxation factors $\gamma$, by using a model where inertial waves are excited via tidal stress in the fluid envelope and dissipated by turbulent friction \citep{Ogilvie2004,Ogilvie2007}, finding similar results to the present work.}

In order to constrain $Q$ for exoplanets, astronomers often refer to studies of Jupiter (see e.g. \citealt{Trilling2000,Bodenheimer2003}). Values for $k_2$ and $Q$ are typically assumed under different considerations as we do not have enough information thus far to ascertain a precise measurement of the deformation of exoplanets, or the efficiency at which these bodies dissipate their internal energy. For instance, \cite{Goldreich1966} \hl{have} constrained the Jovian $Q$ in the range $10^5 - 10^6$, whereas \cite{Yoder1981} propose it is within $6 \times 10^4 - 2 \times 10^6$. These estimates are usually based on tidal evolution modelling of Jovian satellites \citep{Yoder1981,Greenberg1982,Greenberg1989} or on models of the planet's energy dissipation \citep{Goldreich1977,Ogilvie2004,Ogilvie2007}. In this regard we can also achieve further progress via observational constraints \hl{as \cite{Kozai1968,Ray1996} did for the Earth, \cite{Trafton1974} for Neptune, \cite{Dickey1994} for the Moon, \cite{Lainey2009} for Jupiter and Io, \cite{Lainey2012} for Saturn, and additional works by \cite{Husnoo2012} and \cite{Albrecht2012}}.

There are many recent studies about \hl{tidal-dissipated energy} through the viscous dissipation of inertial waves in the solid-fluid structural composition of gas giants \citep*{Ogilvie2004,Ogilvie2007,Ogilvie2013,Guenel2014,Mathis2015a}, as well as the fluid-fluid interiors of stars \citep{Ogilvie2013,Mathis2015b}. \hl{It is} unclear \hl{how they impact the orbital evolution of close-in giant planets}. To address this point, in this work we explore the use of a two-layer body formalism \citep{Ogilvie2013, Guenel2014,Mathis2015a,Mathis2015b} in the study of exoplanet tidal evolution and the evolution of stellar/planetary dissipation properties. In Section \ref{sec:evolution} we introduce a star/planet evolutionary model and in Section \ref{sec:tidal} we include exoplanet tidal evolution. We present the results of this physical-tidal-evolutionary approach, as well as some of \hl{its main limitations and} consequences \hl{for} actual \hl{exoplanets} in Section \ref{sec:results}. Finally, we discuss in Section \ref{sec:conclu} the implications of \hl{this approach} on our understanding of the tidal-induced evolution of a compact extrasolar system.

\section{Bi-layer body evolution: planet and star}
\label{sec:evolution}

\begin{figure}
    \centering
        \includegraphics[scale=0.38]{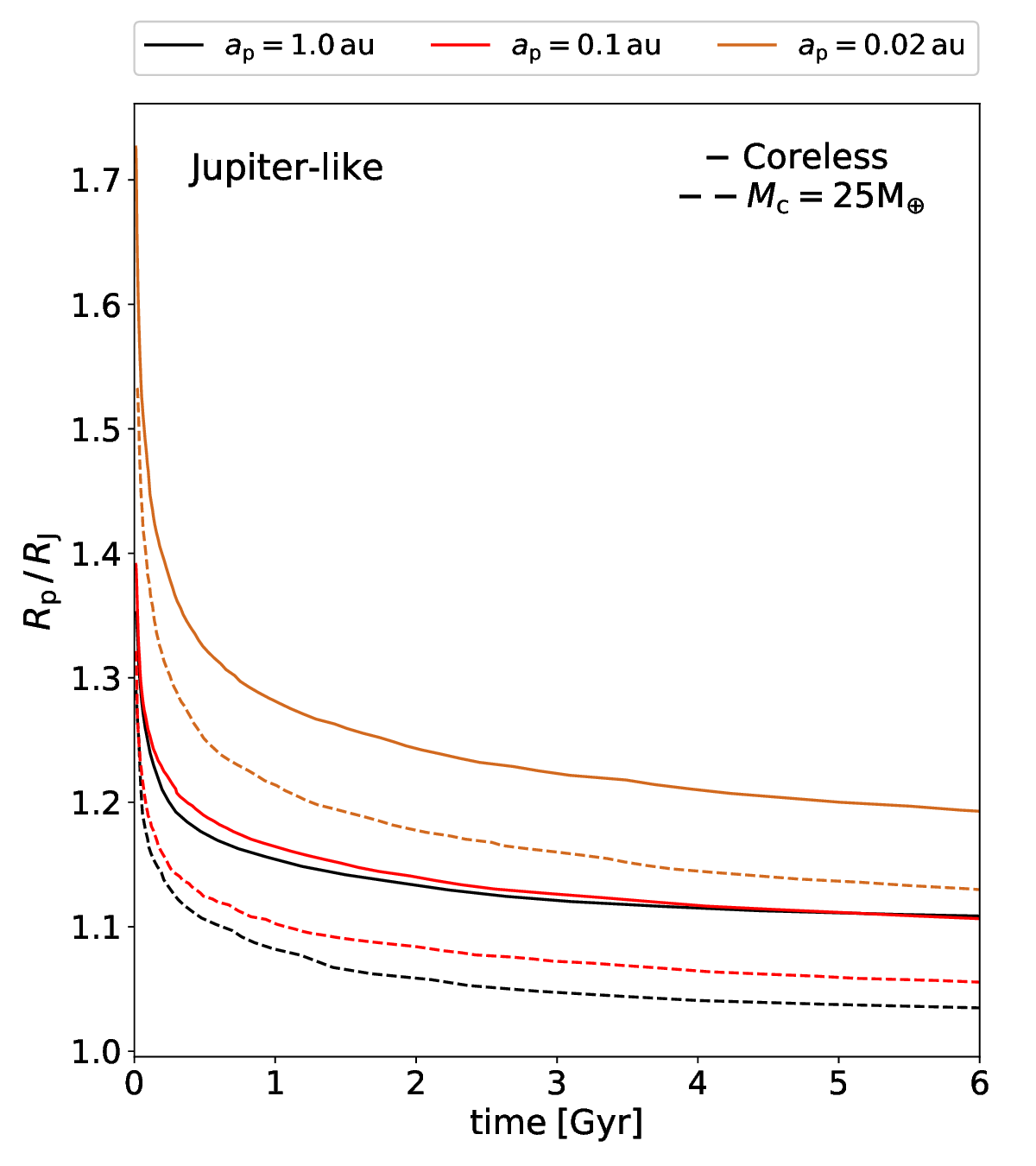}
    \caption{\hl{This plot stands for the deflation of a Jupiter-like planet using the `standard' cooling model proposed by \citep{Fortney2007}. Solid lines represent a planet without core, and dashed lines correspond to a planet with a solid core of 25 $\Mearth$. We have computed for both cases the radius' evolution of a giant planet that is located at different positions from its host star: 0.02 (orange), 0.1 (red) and 1.0 $\AU$ (black). We can notice that for each colour there are two curves (solid and dashed) that start from the same initial radius, but their evolution differs due to their different structural composition, i.e. if the planet has a core or not. For all of the three orbital distances it is evident that a planet with core (dashed lines) has a stronger decrease in size, compared to a coreless planet (solid lines).}}
\label{fig:radiusEvolution}
\end{figure}

In order to test the physical-tidal-evolutionary model for short-period planets proposed in this work, we use the results of \hl{planetary radius contraction} by \cite*{Fortney2007}. These apply for a broad range of planet's masses and compositions, as well as for different planetary semimajor axes (see Fig. \ref{fig:radiusEvolution}). More refined models as that one proposed by \cite*{Miller2009}, \hl{couple} the thermal and tidal evolution \hl{into} the planet's contraction, but they need a different treatment to be implemented in the description of tidal dissipation properties. For our purposes the `standard' cooling model of \cite{Fortney2007} works appropriately.

In the model explored here, we assume that the solid core's mass and radius remain constant during the time-scales concerning to the planetary tidal evolution. When planetary radius $\Rp$, mean planet density $\rhop$ and the planetary rotational rate \hl{$\Op$}\footnote{We point out that the model used here is a simplified two-layer model as in \cite{Remus2012} and \cite{Ogilvie2013}, where the tidally-perturbed body is assumed to be in a state of moderate solid-body rotation (i.e. there is no differential rotation and $\Omega_{\mathrm{core}}$ is neglected). \hl{Thus,} the tidal dissipated energy is generated just from the excited inertial waves of the fluid envelope, \hl{and we omit} any contribution from the inelastic tidal dissipation of the core.} change over time, this has a significant influence affecting gravitationally and mechanically other properties of the planet. With the aim of coupling the evolution of the planet's tidal-related properties, our main effort will be focused on the evolution of what \cite{Ogilvie2013} defines as the `tidal dissipation reservoir' (TDR).
In the classical tidal theory the effect of TDR appears combined in just one quantity (i.e. $Q' = 3Q/2k_2$), where usually $k_2=3/2$ is the (frequency-independent) Love number for a homogeneous fluid body, and $Q$ is the so-called `tidal dissipation quality factor', i.e. the energy in the equilibrium tide divided by the energy dissipated per rotational period \citep{Hansen2010}. However, for the sake of applying the new evolutionary formalism explored in this work for solid-fluid layers of gaseous planets, and fluid-fluid layers of stars, we use the TDR expressed by the ratio $k_2/Q$.

The response of a body to tidal stresses is in general complex and frequency-dependent, as it is in any system forced by oscillation \citep{Efroimsky2012,Remus2012}. The coefficients that measure this are the so-called Love numbers, $k_l^m(\omega)$, and if we assume a low obliquity for the orbit (planets) and a small inclination for the relative axis in the star-planet system (i.e. \hl{co-planar} orbits), the only Love number we cannot neglect is $k_2^2(\omega)$. Physically, $k_2^2(\omega)$ accounts for the tidal-dissipated energy inside a body, as well as the interchange of angular momentum with other bodies by means of rotation and mean motion, e.g. star and planet in this work.

In particular, adopting the formalism of \cite{Ogilvie2013}, if we average on frequency the imaginary part of $k_2^2(\omega)$ we can find an estimation of the TDR $k_2/Q$ as follows,

\beq{eq:Imk2}
\frac{k_2}{Q} =\int_{-\infty}^{+\infty}{\rm Im}[k_2^2(\omega)]\frac{\Der\omega}{\omega}=\int_{-\infty}^{+\infty}\frac{|k_2^2(\omega)|}{Q_2^2(\omega)}\frac{\Der\omega}{\omega}.
\eeq

For our purposes we will adopt the analytical expression of $k_2/Q$ derived from \cite{Ogilvie2013} for planets \citep{Guenel2014} and stars \citep{Mathis2015b} (hereafter, we select either the subscript $\mathrm{p}$ or $\star$  for the planetary or stellar values, respectively),

\beq{eq:k2QFormulap}
\frac{\Kpp}{\Qp} = \frac{100\upi}{63}\epsilon_\mathrm{p}^{2}\frac{\alphap^{5}}{1-\alphap^{5}}\left[1+
	\frac{1-\gammap}{\gammap}\alphap^{3}\right]\left[1+
		\frac{5}{2}\frac{1-\gammap}{\gammap}\alphap^{3}\right]^{-2},
\eeq 

\beq{eq:k2QFormulas}
\begin{split}
    \frac{\Kss}{\Qs} =& \frac{100\upi}{63}\epsilon_\mathrm{\star}^{2}\frac{\alphas^{5}}{1-\alphas^{5}}(1-\gammas^2)(1-\alphas^2)\\&\left(1 + 2\alphas + 3\alphas^2 + \frac{3}{2}\alphas^3\right)^2\left[1+\left(\frac{1-\gammas}{\gammas}\right)\alphas^3\right]\\&\left[1 + \frac{3}{2}\gammas+\frac{5}{2\gammas}\left(1 + \frac{1}{2}\gammas-\frac{3}{2}\gammas^2\right)\alphas^3-\frac{9}{4}(1-\gammas)\alphas^5\right]^{-2}
\end{split}
\eeq

For Eqs. (\ref{eq:k2QFormulap}) and (\ref{eq:k2QFormulas}) $\epsilon_\mathrm{p, \star}^{2}\equiv (\Omega_\mathrm{p,  \star}/\Ocri)^{2}$, where $\Omega_\mathrm{p, \star}$ is the planetary or stellar rotational rate, and $\Ocri=(G M_\mathrm{p,\star}/R_\mathrm{p, \star}^{3})^{1/2}$ the critical value \hl{which provides the dependence on the changes of the body's radius ($R_\mathrm{p,\star}$) and mass ($M_\mathrm{p,\star}$). This critical rotation comes directly from the Kepler's third law, and represents the spin of a point on the surface of each body (planet or star).}

$\alpha_\mathrm{p, \star}$, $\beta_\mathrm{p, \star}$, and $\gamma_\mathrm{p, \star}$ do not have dimensions and are defined as functions of each body's bulk properties,

\beq{eq:k2Qparameters}
\alpha_\mathrm{p, \star}=\frac{\Rc}{R_\mathrm{p,\star}},\; \beta_\mathrm{p, \star}=\frac{\Mc}{M_\mathrm{p,\star}}\;\text{and}\; \gamma_\mathrm{p, \star}=\frac{\rhoe}{\rhoc}=
	\frac{\alpha_\mathrm{p, \star}^{3}(1-\beta_\mathrm{p,  \star})}{\beta_\mathrm{p, \star}(1-\alpha_\mathrm{p, \star}^{3})},
\eeq

\noindent 
where the core's radius, mass, and density are represented by $\Rc$, $\Mc$, and $\rhoc$; $\rhoe$ is the density of the fluid envelope. All of them are for the planet or star, depending on the body we are studying.

The functions $\alpha_\mathrm{p, \star}(t)$, $\beta_\mathrm{p, \star}(t)$, and $\epsilon_\mathrm{p, \star}(t)$ or equivalently $\Rc(t)$, $\Rp(t)$, $\Rstar(t)$, $\Mc(t)$, $\Op(t)$, and $\Os(t)$ need to be estimated in order to apply equations (\ref{eq:k2QFormulap}) and (\ref{eq:k2QFormulas}). That said, we proceed to mark some key points:

    (i) \hl{In general, star and planet are assumed to have a solid-body rotation, and as $\epsilon^2\propto\Omega^2$ then $\epsilon\ll 1$ in equations (\ref{eq:k2QFormulap}) and (\ref{eq:k2QFormulas}). Also, the model proposed on the regime of \cite{Guenel2014} takes into account the Coriolis acceleration, but neglects centrifugal forces. It is worth noticing that for close-in planets, where tidal evolution leads to synchronisation ($\Omega\sim n$), this assumption should be taken carefully.} 

    (ii) We assume that the liquid/solid planet and star core is  already formed, or its evolution is slow enough during most of the envelope contraction so that $\beta_\mathrm{p, \star}(t)=\beta_\mathrm{p, \star}(0)$.
    
    (iii) On the one hand, $\epsilon_\mathrm{p}(t)$ will be computed with the models of planetary evolution (Fig. \ref{fig:radiusEvolution}), and the evolving planet's rotational rate $\Op(t)$ from the tidal interaction with the star (equation \ref{eq:dopdt}). On the other hand, the evolution of $\epsilon_\mathrm{\star}(t)$ results only from the evolution of $\Os(t)$ -- see equation \ref{eq:dosdt} and the last item in this list.
    
    (iv) $\alphap(t)$ will be assumed in an asymptotic value $\alphap(\infty)$, close to that of Solar System planets \citep{Mathis2015a}, and its instantaneous value will be obtained as,

    \beq{eq:alpha}
    \alphap(t)=\alphap(\infty) \frac{\Rp(\infty)}{\Rp(t)} .
    \eeq
    
    \hl{Equation (\ref{eq:alpha}) couples equations (\ref{eq:k2QFormulap}) and (\ref{eq:k2Qparameters}), where $\Rp(t)$ is hereafter computed in all our models from a planet with core located at 0.1 au (see dashed red line in Fig. \ref{fig:radiusEvolution}), and using the cooling contraction model proposed by \citet{Fortney2007}. However, we should keep in mind that planets migrating too close to the parent star could have larger radii and smaller densities, which is an underestimation and overestimation of the present model respectively.}
    
    (v) Finally, the star will be taken with a fixed core's radius. We assume that the star's bulk size does not change significantly during the time-scales of planetary tidal evolution, or $\alphas(t)=\alphas(0)$.

\hl{The validity of the formulae for $\Kpp/\Qp$ (equation \ref{eq:k2QFormulap}) and $\Kss/\Qs$ (equation \ref{eq:k2QFormulas}) adopted in this work is only possible under the previous conditions. It is also very important to stress that although tidal dissipation inside giant planets is still a matter of research and not well understood yet, the central point of the model proposed here is that planetary evolution happens on time-scales similar to those of orbital circularization. This result seems to be pretty robust and model independent.  Also, independent of model is the fact that during the early evolution of the planet $\Rp$, $\Kpp/\Qp$, $\Kss/\Qs$, and the stellar/planetary rotational rates will change, modifying the tidal torque on the evolving planet and its host star.}

We assume that both planet and star have an interior two-layer structural composition (core+envelope): solid-fluid (planet) and fluid-fluid (star). Moreover, in our physical-tidal-evolutionary model we consider only the dissipation of tidal energy in the fluid envelopes caused by the turbulent friction of Coriolis-driven inertial waves, and neglect that dissipated inelastic energy inside the core \citep{Guenel2014}. \hl{For a more precise computing of the mechanisms driving tidal dissipation in gas giants, we must have a better understanding on how rocky and icy layers of planets behave. This means including possible effects of magnetic fields, stable stratification, differential rotation, and in general any non-linear process taking place in planetary layers (e.g. the turbulent friction on fluid envelopes).} 

\hl{We should emphasise that equations (\ref{eq:k2QFormulap}) and (\ref{eq:k2QFormulas}) are frequency-averaged quantities, so the complex dependence on frequency of the dissipative properties in a spherical shell \citep{Ogilvie2007} is being ruled out. As a result, the computed dissipation for a specific frequency might be considerably underestimated or overestimated from its averaged value. In the case of stars, equation (\ref{eq:k2QFormulas}) represents a lower bound of how dissipation is carried out throughout the star, and a more refined model should include the gravito-inertial modes of differential rotation \citep{Ivanov2013}. These play a major role on stellar structure and evolution, but their effects in planetary tidal evolution are not well constrained.}

\section{Tidal-induced Decay}
\label{sec:tidal} 

Planets with considerable orbital eccentricities and located at close-in distances to their parent stars, undergo tidal interaction. This results in an exchange of rotational and orbital angular momentum, and the subsequent circularization, \hl{or orbital decay}, of the planetary orbit. To study this interaction there are different time-dependent quantities we must compute when analysing the evolution of the whole system. For the rotational component, we need to find the instantaneous change of the planetary ($\Op$) and stellar ($\Os$) rotational rates -- by including the evolution on $\Os$ we want to account for the effects of those tides raised by the close-in giant exoplanet on its parent star. 

\hl{Before going any further with the theoretical treatment used in this work on the tidal-induced decay of a close-in giant planet, we should emphasize something important in our model. This is the solid-body approximation we are employing in the numerical study of the stellar and planetary rotational evolution. In a similar framework as that of \cite{Guenel2014, Mathis2015b}, we assume no differential rotation and neglect any contribution coming from the rotation of the solid inner core. For the sake of simplicity, this is the most common avenue and has been explored before in the study of the orbital evolution of short-period planets around slow-rotating stars. It proposes that only the outer convective envelope contributes to the spin-orbit angular momentum exchange \citep{Marcy1997, Dobs2004, Donati2008,Ferraz2008, Jackson2008a}. In terms of the interior structure of gas giants, and the two-layer model proposed here for their rheological properties, the aforementioned approximation is equivalent to neglect the tidal-dissipated energy through the viscoelastic deformation of the solid core in slow rotators. As it was stated in Section \ref{sec:tidal}, we only consider the dissipated energy due to the tidal inertial waves of the fluid envelope.}

\hl{Moreover, the solid core is assumed to be perfectly rigid and homogeneous, and the fluid envelope as homogeneous and incompressible (see \citealt{Ogilvie2013}). In consequence, under these characteristics and the non-differential rotation of each body, we are not accounting for any friction force between the core and the fluid layer. However, between both the solid and fluid layers could exist a friction force, or a gravitational torque, caused by a rotational frequency-lag in the outer layer.}

\hl{That being said,} if $\Mstar\gg\Mp$ \hl{and} the star is aligned and rotates considerably slow (like the Sun), the time variation of $\Os$ and $\Op$ is as follows \citep{Mardling2002},

\beq{eq:dosdt}
\frac{\Der\Os}{\Der t}= \frac{3\npp^4\Mp^2}{\eps\gyrs G}\left(\frac{\Rstar}{\Mstar}\right)^3\frac{\Kss}{\Qs}\left[\hat{e_1}(e)-\hat{e_2}(e)\left(\frac{\Os}{\npp}\right)\right] + \oms,
\eeq

\beq{eq:dopdt}
\frac{\Der\Op}{\Der t}= \frac{3\npp^4\Rp^3}{\epp\gyrp G\Mp}\frac{\Kpp}{\Qp}\left[\hat{e_1}(e)-\hat{e_2}(e)\left(\frac{\Op}{\npp}\right)\right] + \omp
\eeq

where, 

\beq{eq:ecc1}
e_1(e) = \left(1 + \frac{15}{2}e^2 + \frac{45}{8}e^4 + \frac{5}{16}e^6\right) \bigg/ (1-e^2)^6,
\eeq

\beq{eq:ecc2}
e_2(e) = \left(1 + 3e^2 + \frac{3}{8}e^4\right) \bigg/ (1-e^2)^{9/2},
\eeq
and $\epp$/$\eps$ are the fractions of planet/\hl{star} mass contributing to the tidal-induced angular momentum exchange. As the gas giants convection zones are considerable large, their full combination is possible and we can take \hl{$\gyrp\epp\simeq0.2$}, \hl{while} Sun-like stars have shallower surface convection zones and $\gyrs\eps$ is of the order of $\sim10^{-2}$ \hl{\citep{Dobs2004}}. Stellar and planetary TDR are often based on assumed values of the tidal dissipation parameters (i.e. $\Kss, \Kpp, \Qs$ and $\Qp$), but this time are given by equations (\ref{eq:k2QFormulap}) and (\ref{eq:k2QFormulas}). The other quantities are: $G$ the gravitational constant, $\Rp$, $\Mp$, $\gyrp$, $\omp$, and $\Rstar$, $\Mstar$, $\gyrs$, $\oms$ the radius, mass, gyration radius (coefficient of the moment of inertia), and the loss of rotational angular momentum due to the loss of mass via the stellar wind. \hl{Each quantity corresponds to} the planet, \hl{or} the star (subscript $p$ or $\star$), respectively.

The evolution of both rotational rates is coupled, and also determined by  \hl{the planet's} orbital eccentricity $e$. At the same time, we cannot ignore the strong coupling between the tidal evolution of $e$ and that of the planetary mean motion $\npp$ \citep{Jackson2008a}, connected to the evolution of semimajor axis via the Kepler's \hl{third} law. In the case of no planetary inclination ($i_o=0$), these are given by (see \citealt{Ferraz2008,Jackson2008a}):

\beq{eq:dnpdt}
\frac{\Der \npp}{\Der t}=\frac{9}{2}\frac{\Kss}{\Qs}\frac{\Mp}{\Mstar}\left[\frac{\npp^{16}}{(G\Mstar)^{5}}\right]^{1/3}\Rstar^5[1+(23+7\Stide)e^2],
\eeq

\beq{eq:dedt}
\frac{\Der e}{\Der t}=-\frac{3}{2}\frac{\Kss}{\Qs}\frac{\Mp}{\Mstar}\left[\frac{\npp^{13}}{(G\Mstar)^{5}}\right]^{1/3}\Rstar^5(9+7\Stide)e.
\eeq

The only new quantity in the last two equations is $\Stide$,

\beq{eq:tidalratio}
\Stide = \left(\frac{\Kpp}{\Qp}\right)\left(\frac{\Kss}{\Qs}\right)^{-1}\left(\frac{\Mstar}{\Mp}\right)^2\left(\frac{\Rp}{\Rstar}\right)^5
\eeq
which characterises the importance of the relative radii, masses, and planetary/stellar tides. \hl{Equations (\ref{eq:dnpdt}) and (\ref{eq:dedt}) can only be applied under two important considerations of the planetary orbit ($\npp$), and the stellar rotation ($\Os$) as follows: 1) the star is rotating at a rate much slower than the mean motion of the planet  ($\Os\ll\npp$), and 2) the planetary rotation is nearly synchronised with its orbit ($\Op\sim\npp$). Both of these assumptions are valid under the approach considered here, but a more general version of these equations must be used if other scenarios are explored (cf. with those of \citealt{Ferraz2015}), for instance if the stellar rotational period is smaller or comparable to the orbital period ($\Os\la\npp$), .}

\hl{The} set of four differential equations (\ref{eq:dosdt}, \ref{eq:dopdt}, \ref{eq:dnpdt} and \ref{eq:dedt}) \hl{that describe the tidal-decay of a close-in companion around its host star,} is different in the framework of close-in inclined orbits \citep{Barker2009}. However, we constrain our solutions to the co-planar case and let the study of planetary inclination coupled to tidal and interior structure evolution for a future work.

\hl{The major contribution of angular momentum loss in this framework is the interaction between the stellar spin and the planet's orbit. In our solutions we have neglected the contribution that stellar wind makes to the change of rotational angular momentum, called stable conservative systems (i.e. $\oms\simeq\omp\simeq0$). While for the planet is correct assuming $\omp\simeq0$ in post-formation scenarios, $\oms$ may be calculated from the mean rotational velocity for late G and K stars \citep{Skumanich1972} in a similar framework to that explored by \cite{Dobs2004}, who reported that the wind-loss occurs much faster than the orbital synchronisation and subsequent decay. With no stellar wind contribution we are neglecting the magnetic braking on the star, so the approach used in this work does not account for the loss of stellar mass. The validity of the latter assumption lies on the relationship of $\Os$, $\npp$, and the position of the co-rotation radius, which is the distance where these two quantities are in equilibrium and the stellar rotation synchronises with the planet's orbit ($\Os=\npp$).}

\hl{Magnetic braking is more significant for rapidly rotating stars, as they are more affected by the active evolution of their convective zones. In addition, during the stellar lifespan this effect makes that none close-in planet can be considered tidally evolved, as it reduces considerably the tidal friction and causes that $\Os<\npp$ (i.e. planets orbit inside co-rotation). These planets can still be subject to tidal-induced orbital decay. Therefore, magnetic braking on the star increases the time that a planet is stable on its orbit during which the stellar evolution is still present.} 

\hl{The absence of magnetic braking has been well analysed by previous studies (see e.g. \citealt{Counselman1973,Hut1981}), which have shown that $\Os=\npp$ is a stable equilibrium that the system tends to achieve \citep{Barker2009}. Based on the initial state of the system (i.e. if $\Os<\npp$ or $\Os>\npp$), the planet can (under certain conditions of the total angular momentum), or cannot be subject to tidal orbital decay. When \mbox{$\Os<\npp$} and both $\Os$ and $\npp$ never synchronise, the tidal-induced orbital decay and the angular momentum interchange intensifies the gap between the star's rotation and the planet mean motion. Thus, the planet does not stop the inward migration and eventually falls in the stellar surface.}

\hl{The main consequence of including the stellar wind braking is that synchronisation of $\Os$ and $\npp$ is not stable, and the conservation of the total angular momentum is not longer achievable. Additionally, the region where the close-in planet is not tidally affected becomes wider as the star evolves. In other words, tidal torques are not enough to understand the present rotational state of Sun-like stars \citep{Barker2009,Ferraz2015}.} 

\hl{A rapidly rotating G-like star will significantly decrease its rotation rate due to the continuous magnetic braking over a time-scale of some Gyr, like in the case of our Sun. While the total angular momentum decreases, $\Os$ and $\npp$ may temporarily synchronise, but the magnetic braking will eventually drive the system towards a state where $\frac{\Der\Os}{\Der t} > \frac{\Der\npp}{\Der t} > 0$. This means that the star rotation is much slower than the planet's mean motion, unless the latter transfers significant orbital angular momentum to balance out for the braking. Still, the required time-scale for such a scenario is much longer than the star's lifetime, and it is also needed a considerably large amount of angular momentum coming from the planetary orbit to spin up the star's rotation. Assuming that such a scenario is not reached in our model, and that $\Os\ll\npp$ (i.e. the stars are well evolved and far from synchronising their rotation with the planet's orbit), the planet keeps its migration towards the interior of the system and equations (\ref{eq:dnpdt}) and (\ref{eq:dedt}) remain valid.}

\hl{Moreover}, in \hl{many} cases the stellar rotation is much slower than the planet's mean \hl{orbital} motion (i.e. $\Os\ll\npp$). Young \hl{rapidly} rotating stars lose angular momentum because of the star-disk magnetic coupling, and can end with rotation periods $P_\star\ge10$ d in a few Myr \citep*{Tinker2002}. Stellar wind can also slow down the stellar spin via angular momentum interchange \citep{Skumanich1972,Verbunt1981,Ogilvie2007}, keeping the star's \hl{rotation rate much smaller} than the planet's mean \hl{orbital} motion. In order to analyse if the effect of the initial rotational state of the star has a significant consequence on the tidal evolution of exoplanets, in Section \ref{sec:results} we will adopt various stellar rotation periods to compare their main differences.

The rotational $\Op$ (equation \ref{eq:dopdt}) and orbital $\npp$ (equation \ref{eq:dnpdt}) angular momentum contributions of the planet can take different values, depending on the assumed initial states (i.e. $\Oini$ and $\nini$). \hl{In the case of close-in sub-stellar companions, the star-planet angular momentum exchange often leads to the synchronisation of the planetary rotational rate with the planet's mean motion ($\Op\sim\npp$). The time-scales of this process depend on the specific system's physical parameters \citep{Peale1977, Rasio1996}, and the initial states of the stellar rotation and the planet's orbital elements}. For our set of differential equations we adopt a planet \hl{near to a synchronous} state, and neglect any probability that \hl{it} undergoes a non-synchronous resonance between the rotation $\Op$ and mean motion $\npp$ \citep{Winn2005}. 

\hl{Whichever the formation channel of a close-in giant planet is, some planets may migrate well inside from their initial orbital position. The planets used in this work have initial non-circularized (future-circularized) orbits, which are larger than the critical Roche radius in order to trace the whole tidal decay evolution. This critical radius is derived from the delimited zone for an orbiting body (planet) to be disrupted by the gravitational forces of the larger parent body (star) \citep{Roche1849},}

\begin{figure}
    \centering
        \includegraphics[scale=0.39]{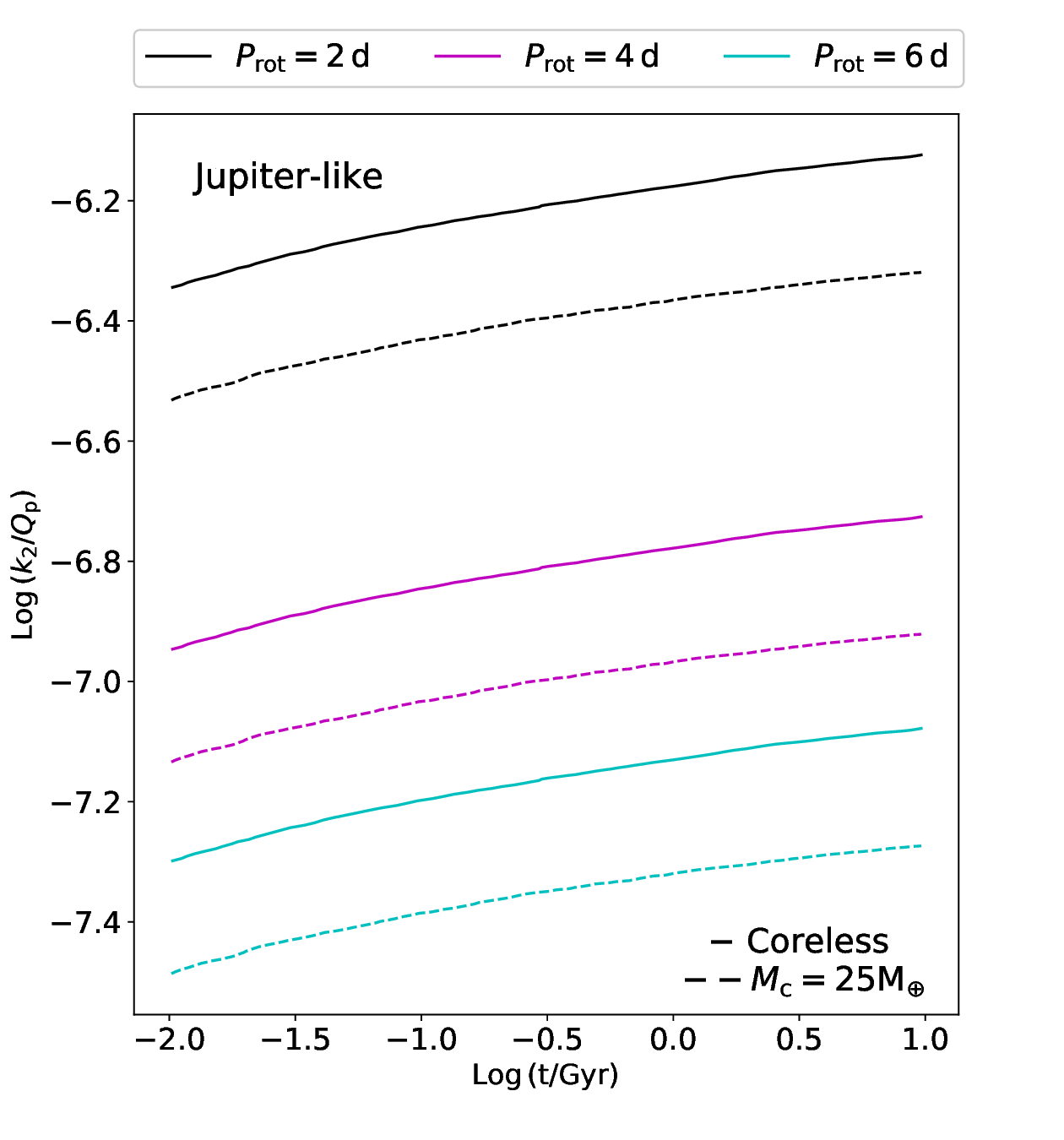}
    \caption{Planetary dissipation function evolution of a Jupiter-like planet for fixed rotational periods using equation (\ref{eq:k2QFormulap}). Solid and dashed lines correspond to planets with no solid core and with a solid core of 25 $\Mearth$, respectively. Evolution has been computed at different planetary rotational periods: 2 (black), 4 (magenta) and 6 d (light blue). \hl{For a planet with core we use the radius' evolution model of a planet located at 0.1 au (dashed red line in Fig. \ref{fig:radiusEvolution}, \citealt{Fortney2007}). In the case of no core, the computing is made by using the evolution of a coreless planet located at 0.1 au (solid red line in Fig. \ref{fig:radiusEvolution}, \citealt{Fortney2007}). For both cases we use equation (\ref{eq:alpha}) to integrate the changes of $\Rp$ into equation (\ref{eq:k2QFormulap}).}}
\label{fig:k2qevolution}
\end{figure}

\begin{figure*}
    \centering
        \includegraphics[scale=0.59]{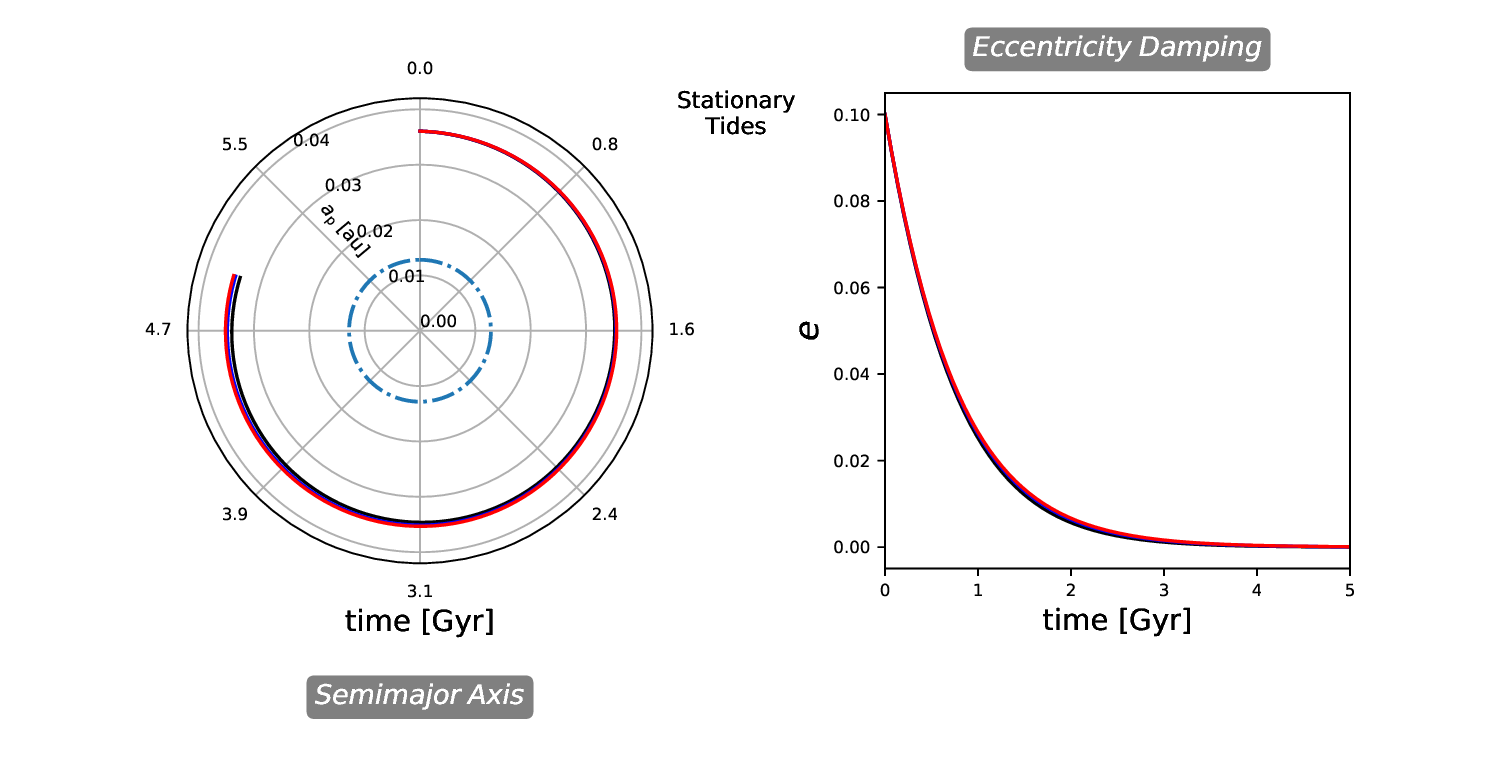}
        \includegraphics[scale=0.59]{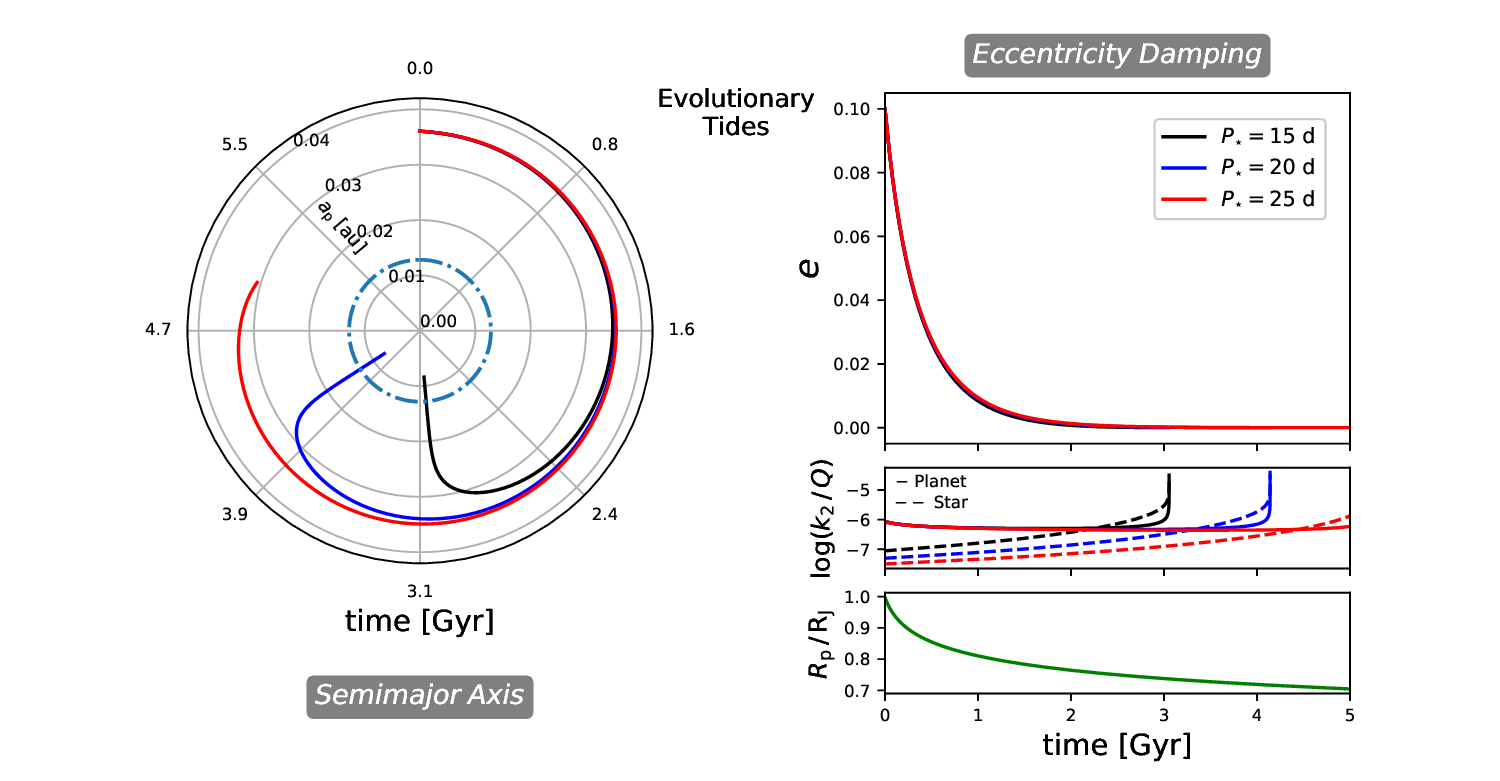}
    \caption{Top-left panel: orbital decay of a hot Jupiter-like planet around a Sun-like star under stationary model of tides. The radial axis illustrates the evolution of the semimajor axis $\apos$, and the polar axis is the time increasing clockwise. Top-right panel: evolution of orbital eccentricity for the corresponding stellar rotational periods. Bottom-left panel: exactly \hl{the} same description of \hl{top-left} panel, but including the evolution of planetary and stellar tides. \hl{Bottom-right panel: eccentricity damping for the tidal evolutionary model (upper), evolution of tidal heat functions (middle), and contraction of planetary radius (lower).} For all of the plots the initial eccentricity is $e_\mathrm{o}=0.1$. \hl{The interior properties of the giant planet are taken with those values of Jupiter, i.e. core mass fraction $\betap=0.02$, and core radius fraction \hl{$\alphap=0.126$}, which are estimated values from \citep{Mathis2015a} according to measurements of Jovian $\koQ$ \citep{Guenel2014}. For the star we use Sun-like values of $\alphas=\betas=0.25$ (\citealt{Garcia2007}).}}
\label{fig:result1}
\end{figure*}

\beq{eq:rtidal}
R_\mathrm{tide} = \eta\left(\frac{\Mstar}{\Mp}\right)^{1/3}\Rp
\eeq
where $\eta$ is a dimensionless factor ranging from 2 to 3, depending on the interior of the planet.
For the sake of concreteness, we take $\eta = 2.7$ based on simulations of the disruption of giant planets \citep*{Guillochon2011}. \hl{Also, our numerical simulations are solely dynamical and no hydrodynamical effects are included, which is an important approximation to keep in mind once a gas giant crosses $R_\mathrm{tide}$. Nevertheless, from left panels in Figs. \ref{fig:result1}, \ref{fig:result2}, and \ref{fig:result3} can be seen that the time necessary to traverse from $R_\mathrm{tide}$ to the surface of the host star is just a very small fraction of the whole tidal evolution. For most cases explored in this work, such time-scales are only of the order of few Myr, and the time-scales of dynamical interactions are still dominant.}

\section{Results}
\label{sec:results}

In this section we coupled the bi-layer model of gas giants and their tidal-related properties (i.e. the second-order Love numbers $k_\mathrm{p, \star}$ and tidal heat functions $Q_\mathrm{p, \star}$ in Section \ref{sec:evolution}, to the tidal decay formalism in Section \ref{sec:tidal}). \hl{In} order to compute the tidal evolution of different \hl{exoplanet} systems, \hl{we use} actual values of the planetary physical and orbital parameters ($\Mp$, $\Rp$, $e$, and $\porb$), with Fig. \ref{fig:distrib} as a reference. We have studied \hl{some specific} cases to analyse the effect \hl{that} different parameters \hl{have} in the tidal evolution of exoplanets. In particular, the initial eccentricity $e_\mathrm{o}$, the stellar rotation period ($P_\star$), and the stellar/planetary radius ($\Rstar$ and $\Rp$) are the ones in which we focus our attention. 

\hl{The interior properties of any giant planet in this work are taken from} those \hl{nominal} values \hl{of the unknown internal bulk properties of} Jupiter -- core mass ($\betap=0.02$) and radius fraction \hl{($\alphap=0.126$)}, \hl{which are expected to be parallel to the close-in hot Jupiters. Such values are} extracted from the \hl{computed core mass and radius fractions} of \cite{Mathis2015a}, who calculated \hl{these properties} for Jupiter, Saturn, Uranus, and Neptune according to actual measurements of $\koQ$ (see Table 1 in that same work). \hl{The initial stellar spin will correspond to three different initial periods ($\prot=$ 15, 20 and 25 d) so we can analyse the influence} that the \hl{star's rotation} has on the tidal evolution of close-in giant planets. 

\hl{By virtue of having reliable values of the stellar core mass ($\alphas$) and radius ($\betas$) fractions, we use Sun-like values for the exoplanet hosting stars; that is, $\alphas=\betas=0.25$ \citep{Garcia2007}. These quantities determine partially} the initial "\underline{T}idal \underline{D}issipation \underline{R}eservoir" (TDR) $\Kss/\Qs$, which is \hl{further} determined by the \hl{stellar rotational rate} via $\epsilon_\star$ (see equation \ref{eq:k2QFormulas}). \hl{We noticed that the study of exoplanet tidal evolution should include the existing connection} between the star's rotation and the tidal dissipated energy per period \hl{in} the star. \hl{This cannot be neglected since the TDR changes considerably for two consecutive stellar spins, see the dashed lines on the bottom-right-middle panel of Fig. \ref{fig:result1}. The same applies for the planetary TDR, where a difference of one order of magnitude can be seen in Fig. \ref{fig:k2qevolution}, depending on the planet's initial rotational period.}

\subsection{Sample study case: Sun-hot Jupiter system}
\label{sec:samplecase}

\hl{We considered two main scenarios for all of the study cases: with and without instantaneous changes on the evolution of tides (called evolutionary and stationary, respectively)}. In the latter scenario the effect of \hl{static tides is included} in the orbital evolution of the planet, but with \hl{constant} nominal values of the TDR \hl{according to the model explained in Section \ref{sec:evolution}, which differ considerably from previous works on tidal evolution} (e.g. \citealt{Dobs2004,Ferraz2008,Jackson2008a,Barker2009,Miller2009,Rodriguez2010}). On the other hand, the evolutionary scenario \hl{computes the instantaneous changes of tides and includes their effect in} the orbital evolution of the planet.

\hl{Firstly}, we numerically integrated our four differential equations for a sample Sun-hot Jupiter system, \hl{where the planet is initially located in a very short-period orbit ($\porb=2.5$ d) with $e_\mathrm{o}=0.1$}. In this numerical scheme, we compute the \hl{stellar/planetary spin, the dynamical} and the tidal evolution by considering \hl{only} single-planet architectures. \hl{For} the Sun-hot Jupiter system we found that eccentricity damping occurs in $\sim4$ Gyr regardless of the stellar spin and the tidal model adopted, either \hl{stationary or evolutionary}. 

As it can be noticed from top-right (stationary) and bottom-right-upper (evolutionary) plots in Fig. \ref{fig:result1}, the difference in the \hl{eccentricity damping} time-scales for each $\prot$ (red, black and blue lines) is almost negligible, and \hl{the circularization occurs} nearly at the same time \hl{for} the three stellar periods. \hl{However, left panels (top and bottom) from this same Fig. \ref{fig:result1} reveal a clear difference between both models of tides, and it is that the inclusion of evolving tides are determinant for the tidal orbital decay of close-in planets, which is in agreement with \cite{Rodriguez2010} who previously found that neglecting the stellar tides retains the planet on its initial orbit without any further significant evolution (see top-left plot in Fig. \ref{fig:result1}).} 

\hl{If we compare our physical-tidal-evolutionary approach to that of \cite{Rodriguez2010} (and similarly to \citealt{Jackson2008a}) who assumed smaller tidal quality factors for the star ($\Qs=1\times10^6$) and the planet ($\Qp=1\times10^5$), we can notice from the bottom-left-middle panel of Fig. \ref{fig:result1} ($k_2/Q$) that in our case these quantities are higher, corresponding to: $\Qs\sim1\times10^7 - 1\times10^{8}$ (dashed lines) and $\Qp\sim1\times10^6$ (solid lines), mainly due to the assumed two-layer frequency-dependent model used to calculate the hot Jupiter's dissipation parameters.} 

The small $\Qs$ and $\Qp$ \hl{values adopted in previous studies result in a strong tidal dissipation from both the star and the planet, which in turn means a faster circularization and tidal decay due to a larger amount of dissipated energy. This means that the planet's orbital angular momentum is swept off in shorter time-scales ( \citealt{Ferraz2008}).} To illustrate this \hll{point}, \hll{a planet around} a star with rotation period of $\prot=15$ d (feasible for a star in the main-sequence, \citealt{Tinker2002}), has a circularization time-scale of $\sim0.1$ \hl{Gyr} \hll{according to} \cite{Rodriguez2010} who assumed \hll{$\Qs\sim1\times10^6$ and $\Qp\sim1\times10^5$}. \hll{However, this time-scale is} $\sim4$ Gyr \hll{when stellar and planetary} tidal functions \hll{are computed through equations (\ref{eq:k2QFormulap}) and (\ref{eq:k2QFormulas}). We found that $\Qs\sim1\times10^{7 - 8}$ and $\Qp\sim1\times10^6$, which is the case for both the stationary (top) and the evolutionary (bottom) scenarios} \hl{in the right panels of Fig. \ref{fig:result1}}. This difference \hll{in circularization time-scales, namely, that of \cite{Rodriguez2010} and this work,} \hl{\hll{depends on how we calculate the initial values} of $\Qs$ and $\Qp$, and on other} physical phenomena \hll{that may take place in the star-planet system.}

\begin{figure*}
    \centering
        \includegraphics[scale=0.59]{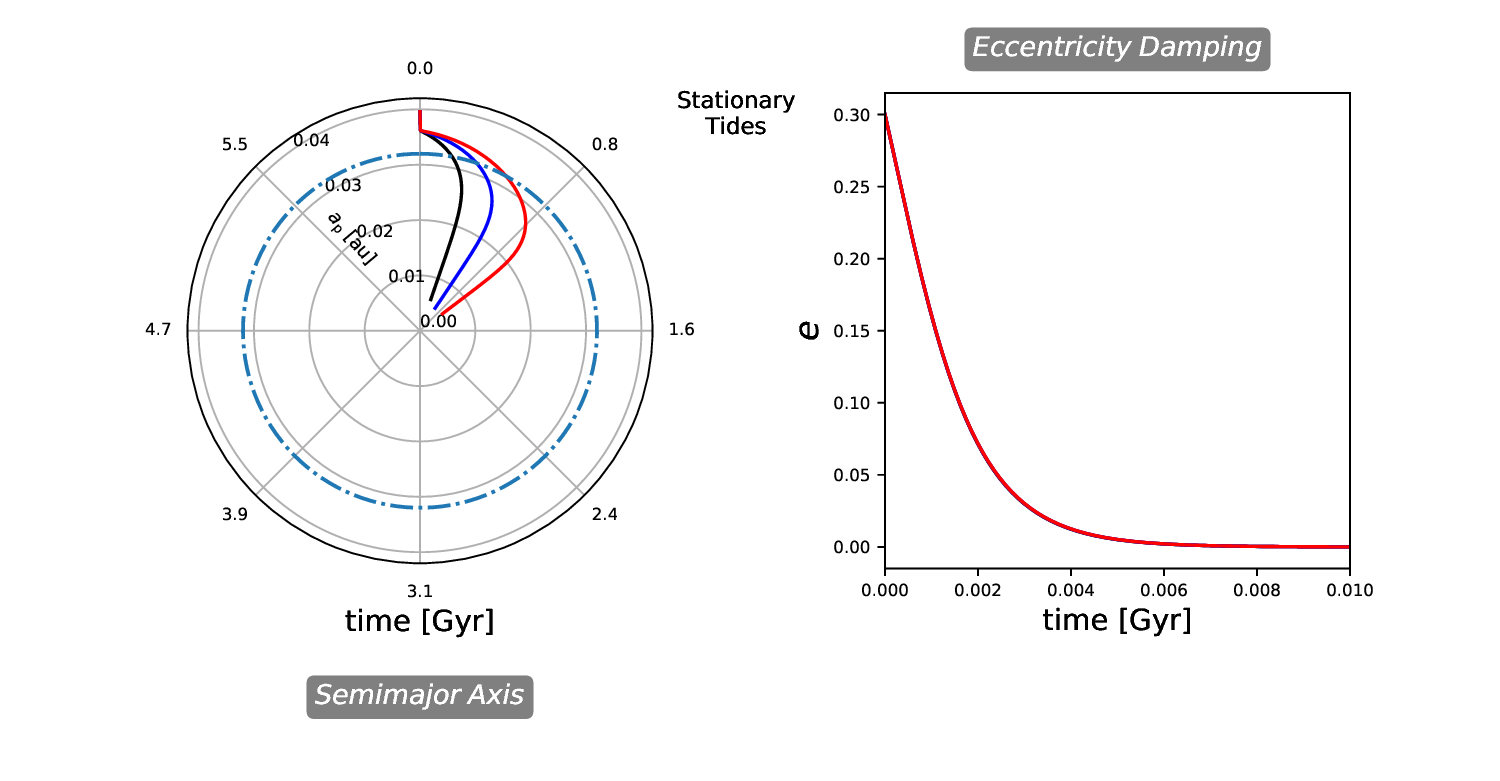}
        \includegraphics[scale=0.59]{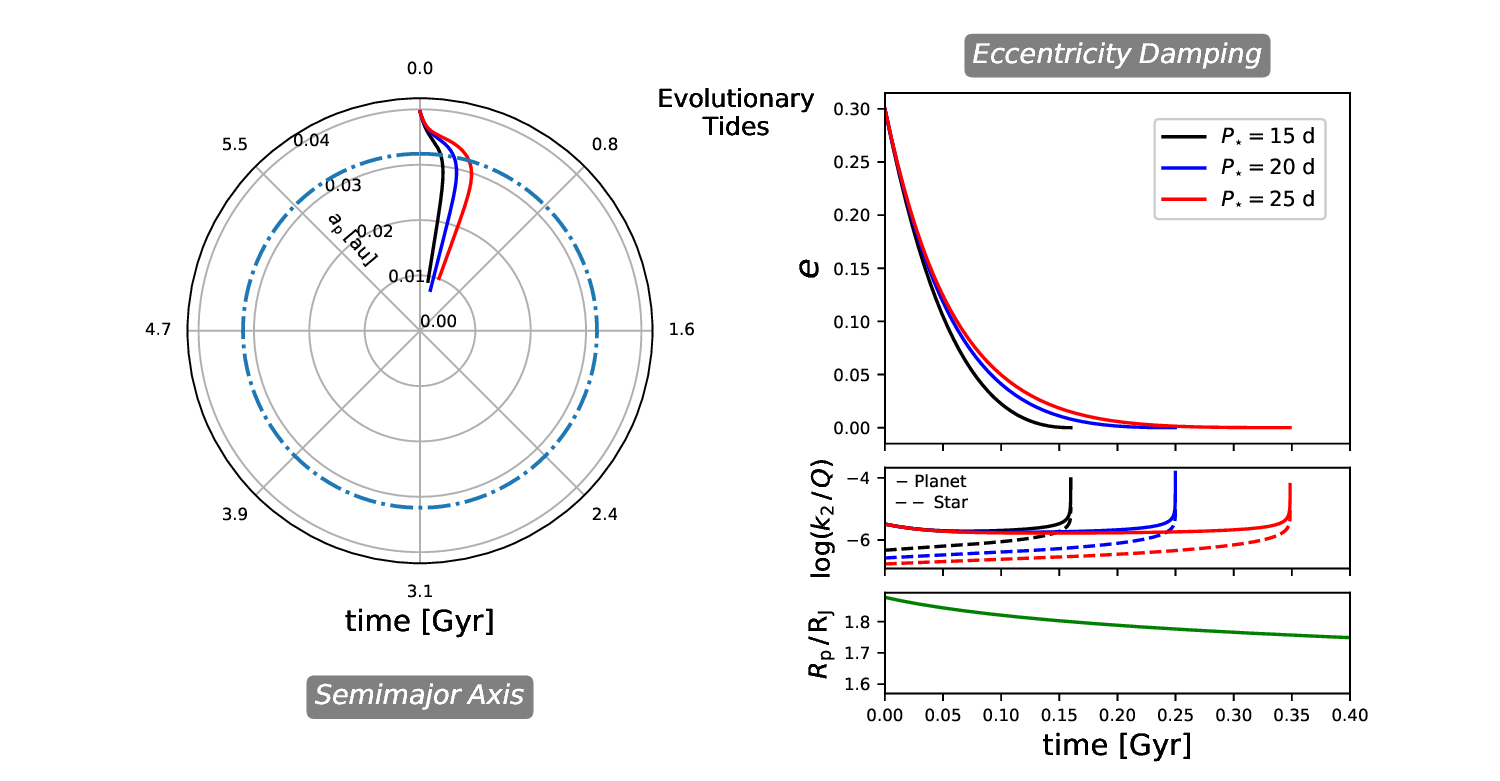}
    \caption{Top-left panel: orbital decay of HAT-P-65 b around its host star under stationary model of tides. The radial axis illustrates the evolution of the semimajor axis $\apos$, and the polar axis is the time increasing clockwise. Top-right panel: evolution of orbital eccentricity for the corresponding stellar rotational periods. Bottom-left panel: exactly \hl{the} same description of \hl{top-left} panel, but including the evolution of planetary and stellar tides. \hl{Bottom-right panel: eccentricity damping for the tidal evolutionary model (upper), evolution of tidal heat functions (middle), and contraction of planetary radius (lower).} For all of the plots the initial eccentricity is $e_\mathrm{o}=0.3$. \hl{The interior properties of the giant planet are taken with those values of Jupiter, i.e. core mass fraction $\betap=0.02$, and core radius fraction \hl{$\alphap=0.126$}, which are estimated values from \citep{Mathis2015a} according to measurements of Jovian $\koQ$ \citep{Guenel2014}. For the star we use Sun-like values of $\alphas=\betas=0.25$ (\citealt{Garcia2007}).}}
\label{fig:result2}
\end{figure*}

\hl{The left panels (top and bottom) of Fig. \ref{fig:result1} \hll{(and also Figs. \ref{fig:result2} and \ref{fig:result3})} help us to understand how the tidal decay occurs for both the stationary and the evolutionary models. They are designed as `orbital clocks', so the polar axis (angles) is time [Gyr] which increases clockwise, and the radial axis stands for the planet's semimajor axis $\apos$ [au]. Each plot has a dashed-dotted line in light blue, which represents the distance $R_\mathrm{tide}$ (i.e. the Roche limit, equation \ref{eq:rtidal}) where the stellar tidal forces start to overpass the gravitational cohesion of the planet, and the latter is ripped out. However, given the short time-scales needed to travel from this distance to the stellar surface, it is possible for the star to gobble down most of the planet before this is disrupted by tides.} 

\hl{Bottom-left panel in Fig. \ref{fig:result1} reveals that} when the evolution of tides via the TDRs ($\Kss/\Qs$ and $\Kpp/\Qp$) is considered into the orbital evolution of the planet, and the effects of the tidal decay are implemented in the modification of the stellar/planetary interior structures (tidal bulges), \hl{the eventual decay of the planetary orbit occurs. In brief, the evolution of tides enhances the orbital decay of short-period giant planets in agreement to previous studies. Strikingly, the time-scales for this to happen are determined by the coupling of $\Kss/\Qs$ and $\Kpp/\Qp$, as can be noticed from bottom-right-middle plot which corresponds to the evolution of both TDRs as functions of time. In Fig. \ref{fig:result1} (bottom-right-lower plot) can also be found the evolution of the planetary radius, which follows the model by \cite{Fortney2007} shown in Fig. \ref{fig:radiusEvolution} (see Section \ref{sec:evolution}).}

For the next two exoplanet systems (Figs. \ref{fig:result2} and \ref{fig:result3}), we have also plotted in light blue \hl{colour} the minimum distance where each planet can hold its gravitational stability against the disruption caused by stellar tides (equation \ref{eq:rtidal}). This distance can be assumed fixed throughout the evolution of the planet, since the only time-dependent term is $\Rp$ and its change is \hl{very small} to move this limit closer to the star.

\subsection{Actual exoplanets: HAT-P-65 b and HATS-32 b}
\label{sec:actualexo}

\begin{figure*}
    \centering
        \includegraphics[scale=0.59]{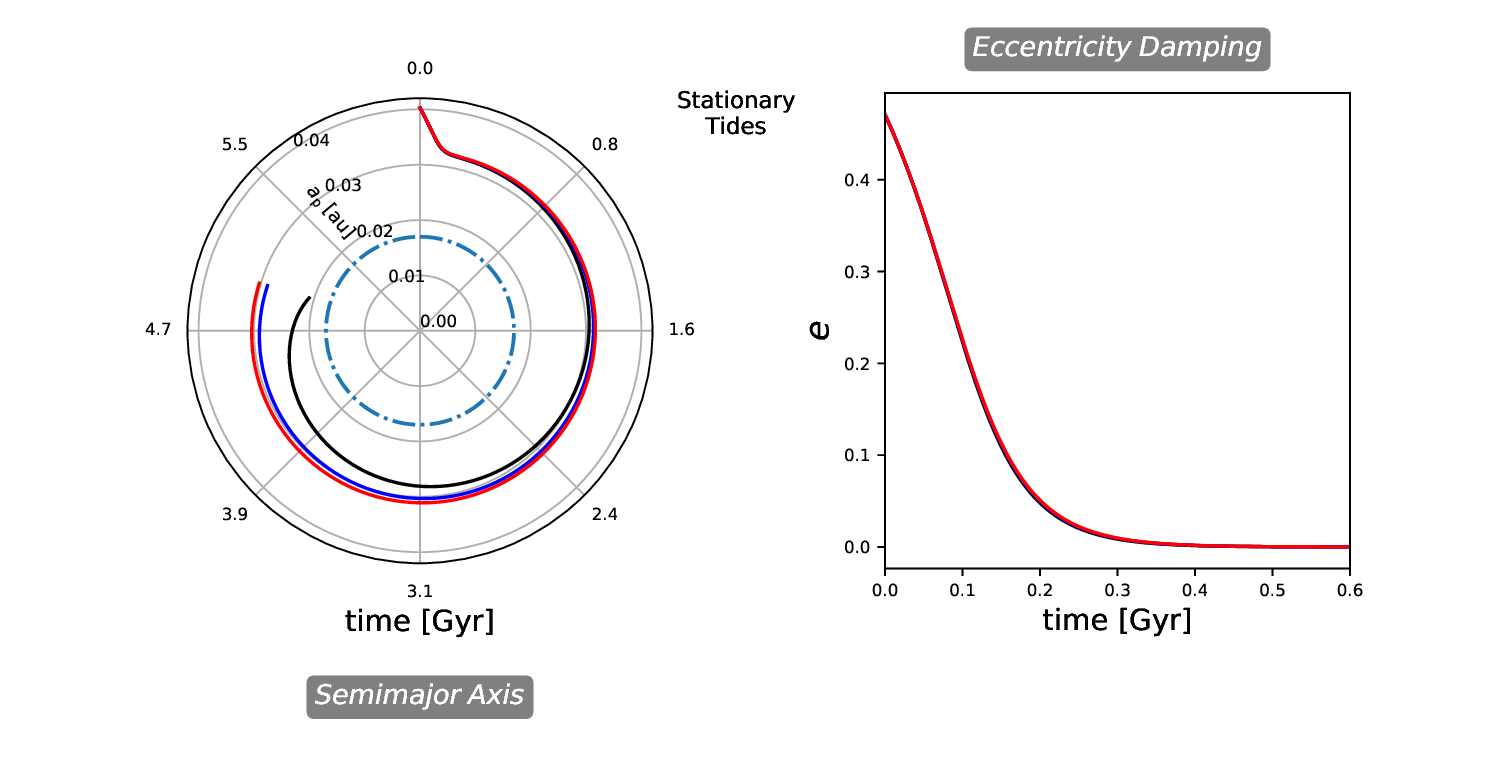}
        \includegraphics[scale=0.59]{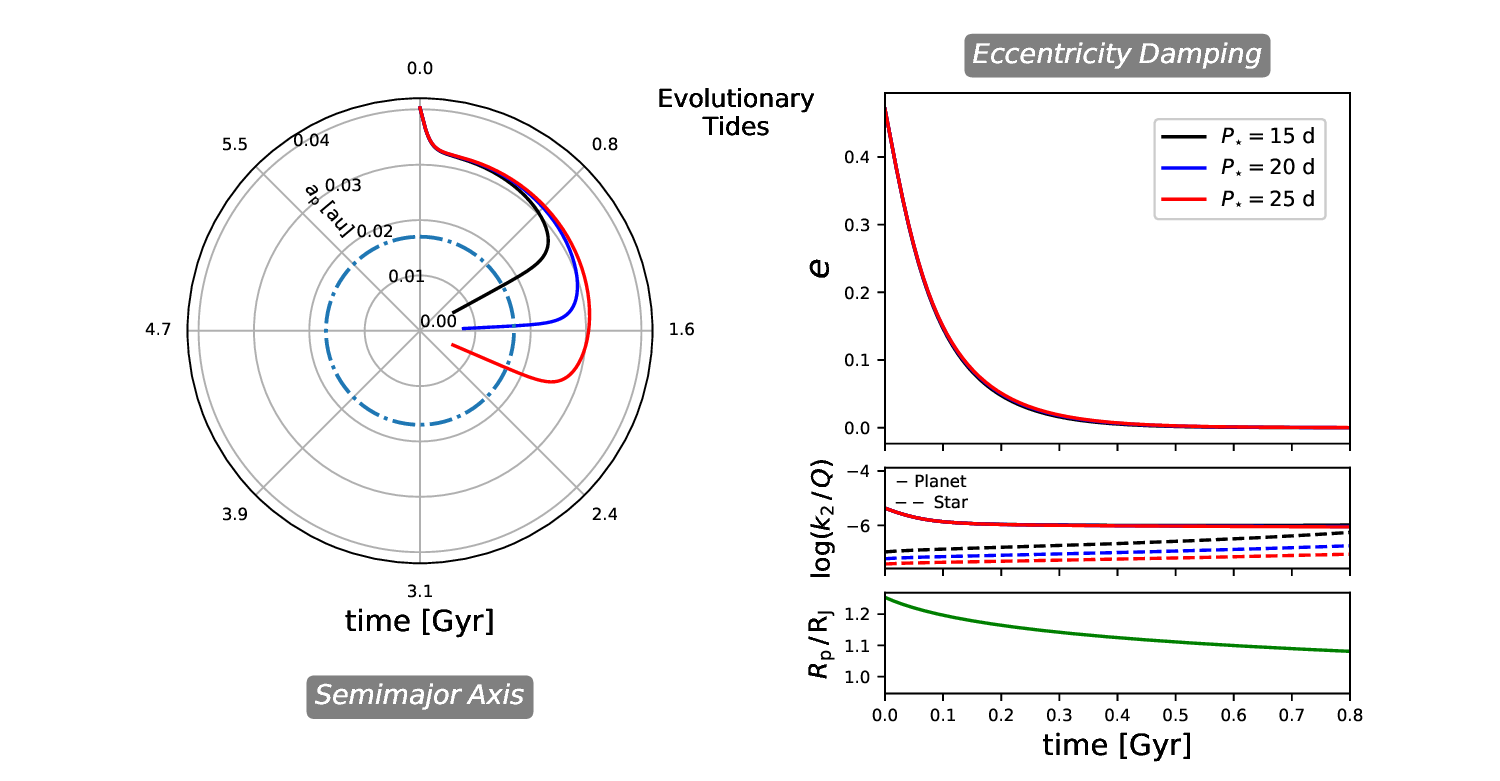}
    \caption{Top-left panel: orbital decay of HATS-32 b around its host star under stationary model of tides. The radial axis illustrates the evolution of the semimajor axis $\apos$, and the polar axis is the time increasing clockwise. Top-right panel: evolution of orbital eccentricity for the corresponding stellar rotational periods. Bottom-left panel: exactly \hl{the} same description of \hl{top-left} panel, but including the evolution of planetary and stellar tides. \hl{Bottom-right panel: eccentricity damping for the tidal evolutionary model (upper), evolution of tidal heat functions (middle), and contraction of planetary radius (lower).} For all of the plots the initial eccentricity is $e_\mathrm{o}=0.471$. \hl{The interior properties of the giant planet are taken with those values of Jupiter, i.e. core mass fraction $\betap=0.02$, and core radius fraction \hl{$\alphap=0.126$}, which are estimated values from \citep{Mathis2015a} according to measurements of Jovian $\koQ$ \citep{Guenel2014}. For the star we use Sun-like values of $\alphas=\betas=0.25$ (\citealt{Garcia2007}).}}
\label{fig:result3}
\end{figure*}

For the exoplanets used in this work, we have taken actual observed data that was extracted\footnote{Using \url{www.exoplanet.eu} for the download and filtering} from \cite{exoeu}. To calculate the tidal evolution of those already confirmed transiting exoplanets, we need some stellar/planetary physical and orbital properties beforehand. The first exoplanet we select is HAT-P-65 b (red square in Fig. \ref{fig:distrib}), with $e_\mathrm{o}=0.3$, $\porb=2.61$ d, $\Rstar=1.86\,\Rsun$, $\Mstar=1.212\,\Msun$, $\Mp=0.527\,\Mjup$, and $\Rp=1.89\,\Rjup$. This is an inflated planet almost twice the size of Jupiter, in a very short-period orbit around a similar (but larger) star to our Sun. \hl{We should note} that the effect of $\porb$ and $\Mstar$ are dependent counterparts, and are directly coupled through the \hl{Kepler's third law}, so we \hl{put our attention} in the effect \hl{that} $e_\mathrm{o}$, $\Rstar$, $\Rp$, and $\Mp$ \hl{have in the orbital evolution}.

\hl{From a comparison between Fig. \ref{fig:result1} ($e_\mathrm{o}=0.1$) and Fig. \ref{fig:result2} ($e_\mathrm{o}=0.3$), we can notice at first that tidal evolution time-scales for both the stationary (top panel) and the evolutionary (bottom panel) models of tides strongly depend on the initial eccentricity $e_\mathrm{o}$, and that} increasing the eccentricity of a planet decreases significantly the eccentricity damping time-scales \hl{in both scenarios. This is the aftermath of the coupled evolution between $\frac{\Der \npp}{\Der t}$ (equation\ref{eq:dnpdt}) and $\frac{\Der e}{\Der t}$ (equation \ref{eq:dedt}), and a general rule for close-in giant planets is: large eccentricities imply fast orbital decay. The latter is the result of a stronger instability when planets are located in too eccentric orbits, which should be the case for many discovered planets ($\sim60$, see Fig. \ref{fig:distrib}) with eccentricities ranging from 0.1 to 0.6, as it is the case of HAT-P-65 b (and also HATS-32 b).} 

\hl{We can see from top-right and \hl{bottom-right-upper} plots in Fig. \ref{fig:result2}, that a planet much larger than the hot Jupiter in Section \ref{sec:samplecase} \hl{but} with similar $\porb$, has shorter circularization time-scales in both models \hl{when compared to the Sun-hot Jupiter sample study case}. This happens because 1) for the planet is also true that $\Kpp/\Qp\propto\Rp^2$, so the `tidal dissipation reservoir' (TDR) increases with the radius, which means that the effect of evolutionary tides is to increase the dissipated energy with time, and 2) the planet started in a more eccentric orbit.}

\hl{Both arguments above are really important and determinant for HAT-P-65 b. For instance, Fig. \ref{fig:result2} shows that contrary to the Sun-hot Jupiter sample case (Fig. \ref{fig:result1}) and HATS-32 b (Fig. \ref{fig:result3}), where the eccentricity damping time-scales are the same for both the stationary and the evolutionary models, these time-scales differ approximately in one order of magnitude each other for HAT-P-65 b. Here, the evolutionary scenario (bottom panel, Fig. \ref{fig:result2}) has a delayed orbital circularization compared to the stationary (top panel, Fig. \ref{fig:result2}).}

A large eccentricity is an overarching effect that shortens the circularization time-scales for both the stationary and the evolutionary models in \mbox{HAT-P-65 b}. This is because \hll{for large eccentricities}, \hll{both} $\frac{\Der \Os}{\Der t}$ and $\frac{\Der \Op}{\Der t}$ go up with \hll{$e_1$} (equation \ref{eq:ecc1}) so that $\Os$ and $\Op$ increase faster. Particularly, the increase in $\Os$ triggers a larger stellar TDR as time goes by since $\Kss/\Qs\propto\Os^2$ (equation \ref{eq:k2QFormulas}). Then, as $\frac{\Der e}{\Der t}\propto-\Kss/\Qs$, the decreasing rate of the eccentricity becomes more negative, and thereby $e$ decreases slower when compared to the stationary case, where $\Kss/\Qs$ was fixed and did not evolve. 

In addition, as we know that $\Kpp/\Qp\propto\Rp^2$ (equation \ref{eq:k2QFormulap}), \hll{and HAT-P-65 b is a very inflated planet, this affects significantly the computing of its tidal properties. In the evolutionary model (bottom panel in Fig. \ref{fig:result2}), as $\Rp$ decreases, $\alphap$ increases, so $\Kpp/\Qp$ goes up (see equations \ref{eq:k2QFormulap} and \ref{eq:k2Qparameters}). This causes that $\Stide$ (equation \ref{eq:tidalratio}) increases and again $\frac{\Der e}{\Der t}\propto\Stide$ becomes more negative (i.e. $e$ decreases slower). Thus, the eccentricity damping is delayed respect to the stationary model when $\Rp$ is very large, in contrast to a more contracted planet like the hot Jupiter in Fig. \ref{fig:result1}, or HATS-32 b in Fig. \ref{fig:result3}, where both models have nearly the same time-scales.}

For HAT-P-65 b the high increase in $\Kss/\Qs$ marks the pace of $\Kpp/\Qp$, as $\frac{\Der \npp}{\Der t}\propto\Kss/\Qs$ (equation \ref{eq:dnpdt}), so $\npp$ \hll{is increasing faster} and so does $\frac{\Der \Op}{\Der t}\propto\npp^4$ (see equation \ref{eq:dopdt}). Therefore, as $\Kpp/\Qp\propto\Op^2$ (equation \ref{eq:k2QFormulap}) the planetary TDR is increasing via $\Op$ and $\Rp$ (i.e. double effect), and the coupling with $\Kss/\Qs$ occurs exactly when the orbital circularizaion takes place (see bottom-right-middle plot in Fig. \ref{fig:result2}). 

Additionally, HAT-P-65 is a star almost twice the size ($\Rstar$) of our Sun, and as expected $\Kss/\Qs$ has a strong dependence on size via the critical rotational rate ($\Ocri$) and the radius core fraction ($\alphas$). As the stellar radius increases, the value of $\alphas$ goes down as $1/\Rstar$; and $\epsilon_\star^2$ goes up with $\Rstar^3$. For small values of $\alphas$, $\Kss/\Qs\propto\epsilon_\star^2\alphas^5\propto\Rstar^2$ (see equation \ref{eq:k2QFormulas}), so a larger stellar size stabilises $\Kss/\Qs$ towards a fixed value -- large stellar radii shift the system towards a stationary state. \hl{Then, the aforementioned different time-scales between both models for HAT-P-65 b (see top-right and bottom-right-upper plots in Fig. \ref{fig:result2}) are due to the combined effect of $\Op$, $\Rp$, and $\Rstar$.}

For all of the study cases in this work, the evolutionary scenario also includes the evolution of $\Rp$ (see bottom-right-lower plot in Figs. \ref{fig:result1}, \ref{fig:result2}, and \ref{fig:result3}), \hll{which induces variations in both tidal dissipation parameters $\Kpp/\Qp$ (equation \ref{eq:k2QFormulap}), and $\Kss/\Qs$ (equation \ref{eq:k2QFormulas}). This is one of the reasons for the different effect that evolutionary tides produce in planets like HAT-P-65 b (more inflated, bottom-right-lower panel in Fig. \ref{fig:result2}), and the sample hot Jupiter (more contracted, bottom-right-lower panel in Fig. \ref{fig:result1}). If we compare Figs. \ref{fig:result1} and \ref{fig:result2}, we can further notice that the variation of $\Kss/\Qs$ is similar (bottom-right-middle panel in both figures), but this is the effect due to the large $\Rstar$ of HAT-P-65, as explained in the previous paragraph. The only difference is that for HAT-P-65 b the energy dissipation occurs faster (in both models) due to its high eccentricity, but is delayed when tides evolve because of the large $\Rp$.}

Top-left (stationary) and bottom-left (evolutionary) panels in Fig. \ref{fig:result2} reveal that orbital decay occurs for both scenarios, and that including the evolution of stellar/planetary tides does not change considerably the final fate of the planet. \hl{However, for the tidal evolutionary model, the time that planets remain outside of the minimum self-gravitation distance ($R_\mathrm{tide}$) is shorter than in the stationary scenario, due mainly to the asymptotic decrease of the planet's semimajor axis in the latter.} From all the studied cases and other explored scenarios not included in this work, we could note that the time evolution of tides during the orbital decay of giant exoplanets is important and, hence, has an influence when calculating the eccentricity damping time-scales of short-period planets.

We also studied HATS-32 b (Fig. \ref{fig:result3}) with $e_\mathrm{o}=0.471$, $\porb=2.813$ d, $\Rstar=1.097\,\Rsun$, $\Mstar=1.099\,\Msun$, $\Mp=0.92\,\Mjup$, and $\Rp=1.249\,\Rjup$. This system is very similar to our Sun-hot Jupiter sample case, but with the difference that the planet is less massive and \hl{has a radius slightly greater} than Jupiter. \hl{Bottom-right-middle plot in Fig. \ref{fig:result3} shows that the coupling between $\Kss/\Qs$ and $\Kpp/\Qp$ does not occur in the evolutionary model for \mbox{HATS-32 b} when the circularization is achieved. This is contrary to the Sun-hot Jupiter sample case and HAT-P-65 b, where the circularization took place at the same time that the orbital decay started to be evident (see Figs. \ref{fig:result1} and \ref{fig:result2}). In terms of the time-scales for both the stationary and the evolutionary tidal models in HATS-32 b, the eccentricity damping in this case (comparing top-right and bottom-right-upper plots in Fig.  \ref{fig:result3}) does not differ each other. In this case the only dominant effect is that of the large eccentricity, but the other parameters are very similar to the Sun-hot Jupiter system, which are counterbalanced via  $\Stide$ in equation (\ref{eq:tidalratio}).}

\hl{Finally, a general result for all of the study cases is that orbital decay (left panels in Figs. \ref{fig:result1}, \ref{fig:result2} and \ref{fig:result3}) occurs faster in the evolutionary model rather than in the stationary. The  explanation relies on the underlying physical principle behind the tidal evolution: a planet with a non-negligible eccentricity interacts more closely with the host star in the periastron, so the exchange of spin-orbit angular momentum is enhanced and the star's rotation is accelerated. This increases the efficiency at which the energy is dissipated from the star and the planet, so $\frac{\Der \npp}{\Der t}$ increases and the planet's semimajor axis is becoming closer to the star. This is a result that arises because the stellar and planetary dissipation parameters (i.e. the tidal heat functions $\Kss/\Qp$ and $\Kpp/\Qp$) are directly proportional to the rotational rates of each body in the proposed model, so the exchange of orbital angular momentum is further increased.}

\section{Summary and Discussion}
\label{sec:conclu}

\hl{The interaction and tidal deformation of two (or more) extended bodies, which have close-in relative orbits, is beyond of dispute. Such interactions lead to variations in the orbit of the smaller sub-stellar companion, and for considerable planet sizes, it also has an effect on the interior structural composition of both bodies. This is the case for most of the star-planet extrasolar systems where many discovered planets are comparable to the size of Jupiter and dwell in orbits extremely close to their parent star. Therefore, the study of such planets is necessary to assess the configuration of the whole system where they belong to.} 

\hl{For short-period planets there is an important component that determines most of their recent past and future history, namely the so-called tidal friction. The fact that both bodies are in a high proximity produces an exchange of rotational angular momentum, and this exchange determines their tidal interaction. In order to study the change on the planet's orbital elements and the rotation of both the star and the planet, we assumed here the most common case for exoplanets, i.e. the scenario when the star rotates at a much slower rate than the mean orbital motion of its companion ($\Os\ll\npp$).}

\hl{The assumption above is correct for well evolved stars (see e.g. \citealt{Donati2008}), where the synchronisation of both the stellar and planetary rotational rates is unlikely to occur. Furthermore, having $\Os\ll\npp$ neglects the contribution of the stellar wind in the total angular momentum of the system, which is usually known as magnetic braking. This is particularly important to study the evolution of planets from formation to their current state, when the stellar spin was more active and the tidal dissipation through the stellar wind was significantly high. However, it is feasibly that most observed hot Jupiters are orbiting around stars that have already spun down, so magnetic braking can be presumably ignored for computing their future tidal evolution. If $\Os\ll\npp$ is not satisfied, stellar wind braking should be included not to alter the actual result of planetary tidal evolution \citep{Ferraz2015}.}

We have shown \hl{here} that for short-period planets, their physical evolution plays a key role on the tidal interaction strength with their host star. More importantly, the eccentricity damping (a.k.a. circularization) and the subsequent orbital decay are also affected. This means that the orbital history of a planet is modified at the same time that its bulk and interior properties are \hl{evolving}. In addition, as tides are also produced by the planet on the star \citep{Rasio1996,Hut1981}, we coupled the stellar tidal evolution to the tidal and orbital evolution of the planet. This because both types of tides are important to explain the evolution of extrasolar systems (see e.g. \citealt{Ferraz2008,Rodriguez2010}).

We know that gaseous planets change significantly during the early stages of evolution. Particularly, their size ($\Rp$) and tidal-related properties (gyration radius $\gyrp$, Love number $\Kpp$, dissipated energy $\Qp$, etc.) are prone to change in time-scales of the order of hundreds of Myrs. This `inflation/deflation' derives mostly from their atmosphere contraction, and despite this has been included in previous models for the tidal evolution of exoplanets (e.g. \citealt{Jackson2008a,Miller2009}), the net instantaneous variation of tidal dissipation parameters has not been considered. Stellar and planetary interior properties are often assumed as static values and, therefore, the dissipated energy does not change over time along with the exchange of angular momentum. Such assumptions are often correct since how the energy dissipation occurs in the interior of exoplanets is \hl{poorly} unknown, and the complex tidal processes taking place in a star-planet system are far to be totally understood.

\hl{Previous works have assumed (see e.g. \citealt{Jackson2008a}) a fixed tidal quality factor for stars ($\Qs$), as well as the same $\Qp$ for all short-period exoplanets. The latter is based on assumptions of their interior structures, which are expected to resemble that of Jupiter. This approach has been widely used since the tidal properties of exoplanets are highly uncertain, but there is clearly a shortage of information regarding to their interior structures. Therefore, more independent studies are needed to shed some light on these `unknown premises' which will allow us to know with better detail the history of extrasolar systems.} Here, we implemented instantaneous changes in the tidal dissipation reservoir (TDR) as a result of a modified interior composition. Such changes occur at the same time than the rotational rates are evolving, i.e. due to dissipated energy and planet's contraction, $\Kss/\Qs$ changes with $\Os$ and $\Kpp/\Qp$ with $\Op$ and $\Rp$.

For a Sun-hot Jupiter \hl{sample} system we found that the circularization time-scales are considerably larger when \hl{tidal dissipation paramaters are calculated in terms of the stellar and planetary interior structural compositions. This is caused by an overestimation in previous studies of the initial efficiency at which bodies dissipate their energy}. \hl{However}, the overall result of the model proposed here \hl{has been confirmed} by previous models in the literature, which propose that tides have to evolve for the planet to experience orbital decay \citep{Ferraz2008,Jackson2008a,Rodriguez2010}.

The radius and mass of the planet's core are two important and uncertain quantities in the proposed model. Without an accurate assessment of the whole planetary orbital evolution we can only use estimates by following those known compositions of giant companions in the Solar System (e.g. as we did using Jupiter's values from \citealt{Mathis2015a}). Moreover, effects like tidal heating further difficult the estimation of core's properties in large planets, and the interactions with other bodies in the system \citep{Rodriguez2011} significantly affect the physical-tidal-evolutionary model of this work. Here we have used single-planet systems in order to keep the model simple. We have estimated that the tidal-related properties of the system represented by $\Kss/\Qs$ and $\Kpp/\Qp$ can vary from one to two orders of magnitude \hl{depending on the rotational rates}. This might also hinder how the star-planet tidal interaction is carried out, \hl{as well as} the corresponding time-scale \hl{of the planet's orbital circularization.}

Studying how tides affect close-in exoplanets is important to trace their past and future. If we know how the orbital parameters of planets respond to tidal stresses, and how the interior structure of a star/planet is modified by those orbital changes, we could determine how those detected systems have evolved towards compact configurations. In addition, the circularization time-scale for those systems which still show orbital eccentricity might be found. \hl{For the time being, this work can be taken as a complement of previous exoplanet tidal evolutionary models, accounting for} tidal changes in two-layer close-in gas giants. This reveals that non-static tides should have a significant impact on planetary orbital evolution. Moreover, short-period planets are the most abundant exoplanets thus far, and this model is a feasible explanation to those planets we know are not currently circularized. \hl{In this pursuit, the outcomes presented here} may be applied to constrain the evolution of planets, and \hl{contribute towards a better understanding of how tidal interactions affect the dynamics of extrasolar systems. This will be further improved} with new \hl{exoplanet} discoveries from ground- and space-based telescopes.

\section*{Acknowledgements}

The authors greatly appreciate the comments and corrections of the referee, Dr. Adri\'an Rodr\'iguez Colucci, which significantly helped not only to improve the manuscript, but also to understand more about this research. Jaime A. Alvarado-Montes acknowledges funding support from Macquarie University through the Macquarie University Research Excellence Scholarship (`iMQRES MRES'). Jaime A. Alvarado-Montes and Carolina Garc\'ia-Carmona gratefully thank Lee Spitler for his thoughtful remarks in the final design of the manuscript. This research has made use of NASA's Astrophysics Data System.




\bibliographystyle{mnras}







\bsp	
\label{lastpage}
\end{document}